\documentclass[preprint,amsmath,amssymb,aps,showkeys,showpacs]{revtex4}
\usepackage[english]{babel}
\usepackage{graphicx}
\usepackage{graphics}
\usepackage{amsmath}
\usepackage{dcolumn}
\usepackage{amssymb}
\usepackage{bm}

\begin{document}
\title{Coulomb-type interaction under Lorentz symmetry breaking effects}
\author{R. L. L. Vit\'oria}
\affiliation{Departamento de F\'isica, Universidade Federal da Para\'iba, Caixa Postal 5008, 58051-900, Jo\~ao Pessoa-PB, Brazil.} 

\author{H. Belich} 
\email{belichjr@gmail.com}
\affiliation{Departamento de F\'isica e Qu\'imica, Universidade Federal do Esp\'irito Santo, Av. Fernando Ferrari, 514, Goiabeiras, 29060-900, Vit\'oria, ES, Brazil.}

\author{K. Bakke}
\email{kbakke@fisica.ufpb.br} 
\affiliation{Departamento de F\'isica, Universidade Federal da Para\'iba, Caixa Postal 5008, 58051-900, Jo\~ao Pessoa-PB, Brazil.}

\begin{abstract}

Based on models of confinement of quarks, we analyse a relativistic scalar particle subject to a scalar potential proportional to the inverse of the radial distance and under the effects of the violation of the Lorentz symmetry. We show that the effects of the Lorentz symmetry breaking can induced a harmonic-type potential. Then, we solve the Klein-Gordon equation analytically and discuss the influence of the background of the violation of the Lorentz symmetry on the relativistic energy levels. 

\end{abstract}


\maketitle

\section{Introduction}

The search for physics beyond the Standard Model (SM) has been increased with the need for understanding new phenomena, such as, the unbalance between matter-antimatter and the dark matter. With respect to the dark matter, it is expected that the dark sector can interact with the visible sector, and thus, it can induce the detection of a weak fifth force in such a way that it can be investigated in decays of an excited state of $^{8}\mathrm{Be}$ \cite{irvin}. 

From the studies of quantum chromodynamics (QCD), the size of the proton radius can be estimated through the quark interaction mediated by virtual gluons. Further, experiments with electrons have shown that the value of the size of the proton radius is in agreement with that yielded by the existing theories. However, recent researchers have considered a muon in orbit around a proton and shown that the radius of the proton is different \cite{sci}. It is worth pointing out that these facts occurred just when SM goes through its final test: the detection of the Higgs boson \cite{higgs} at the LHC. Hence, it is necessary to clarify a fundamental question: the Higgs mass. Therefore, these phenomena need a theory that goes beyond SM. In the last decades, the search for a more fundamental theory has given rise to proposal that became known as the Standard Model Extension (SME) \cite{sam,col,coll2}. An interesting point of SME is that there exist terms that violate the Lorentz symmetry by imposing at least one privileged direction in the spacetime. In recent decades, studies of the violation of the Lorentz symmetry have been made in several branches of physics \cite{h1,h2,e1,e2,e3,e4,e5,e6,e7,rasb1,w,tensor1,tensor2,geom1,geom2,geom3,bb2,bb4,book,l1,g1,g2,g3,baeta,bras,gravity,gravity2,louzada,jackiw}.

In this paper, we consider a relativistic scalar particle subject to a scalar potential proportional to the inverse of the radial distance. We also consider this particle to be under the effects of the violation of the Lorentz symmetry. The violation of the Lorentz symmetry is established by a tensor field. Then, we chose a particular background of the violation of the Lorentz symmetry that yields a harmonic-type potential. Thereby, we show that the Klein-Gordon equation can be solved analytically, and then, discuss the influence of the background of the violation of the Lorentz symmetry on the relativistic energy levels.

The structure of this paper is as follows: in section II, we introduce a scalar potential by modifying the mass term of the relativistic equation. We also introduce the background of the Lorentz symmetry violation defined by a tensor field that governs the Lorentz symmetry violation out of SME. Thus, we consider a background of the violation of the Lorentz symmetry yielded by a radial electric field and a non-null component of the Lorentz symmetry breaking tensor and solve the Klein-Gordon equation analytically; in section III, we present our conclusions.

\section{Relativistic effects}

In this section, we consider a relativistic scalar particle subject to a scalar potential proportional to the inverse of the radial distance, and thus, we investigate the effects of a harmonic-type potential produced by an anisotropic environment generated by a violating term of the Lorentz symmetry on this system. We deal with a model that goes for energy scales beyond the Standard Model. In recent years, two of us  \cite{bb19,bb20} have studied the relativistic quantum dynamics of a scalar particle in a background that breaks the Lorentz symmetry based on the model investigated in Refs. \cite{col,coll2,kost2,kost3,louzada}, where it is introduced a nonminimal coupling into the Klein-Gordon equation given by $\hat{p}^{\mu}\hat{p}_{\mu}\rightarrow \hat{p}^{\mu}\hat{p}_{\mu}+\frac{g}{4}\,\left(K_{F}\right)_{\mu\nu\alpha\beta}\,F^{\mu\nu}\left(x\right)\,F^{\alpha\beta}\left(x\right)$, where $g$ is a constant, $F^{\mu\nu}\left(x\right)$ is the electromagnetic tensor and $\left(K_{F}\right)_{\mu\nu\alpha\beta}$ is the tensor that governs the Lorentz symmetry violation out of SME \cite{col,coll2,kost2,baeta}. From the properties of the tensor $\left(K_{F}\right)_{\mu\nu\alpha\beta}$ \cite{louzada}, it is well-known that it can be decomposed into $3\times3$ matrices that give its parity-even sector ($\left(\kappa_{DE}\right)_{jk}$ and $\left(\kappa_{HB}\right)_{jk}$) and its parity-odd sector ($\left(\kappa_{DB}\right)_{jk}=-\left(\kappa_{HE}\right)_{kj}$) \footnote{The matrices of the parity-even sector are defined as $\left(\kappa_{DE}\right)_{jk}=-2\left(K_{F}\right)_{0j0k}$ and $\left(\kappa_{HB}\right)_{jk}=\frac{1}{2}\epsilon_{jpq}\,\epsilon_{klm}\left(K_{F}\right)^{pqlm}$ and are symmetric. On the other hand, the matrices of the parity-odd sector are defined as $\left(\kappa_{DB}\right)_{jk}=-\left(\kappa_{HE}\right)_{kj}=\epsilon_{kpq}\left(K_{F}\right)^{0jpq}$ and have no symmetry.}. Thereby, the Klein-Gordon equation can be written in the form \cite{louzada,bb20,vbb}:
\begin{eqnarray}
\hat{p}^{\mu}\hat{p}_{\mu}\Phi-\frac{g}{2}\left(\kappa_{DE}\right)_{i\,j}\,E^{i}\,E^{j}\,\Phi+\frac{g}{2}\,\left(\kappa_{HB}\right)_{j\,k}\,B^{i}\,B^{j}\,\Phi-g\left(\kappa_{DB}\right)_{j\,k}\,E^{i}\,B^{j}\,\Phi=m^{2}\,\Phi.
\label{1.1}
\end{eqnarray}

On the other hand, as discussed in Ref. \cite{greiner}, a scalar potential can be introduced into the Klein-Gordon equation through the modification of the mass term: $m\rightarrow m+S\left(\vec{r},\,t\right)$, where $m$ is a constant that corresponds to the mass of the free particle and $S\left(\vec{r},\,t\right)$ is a scalar potential. It is worth mentioning that, by considering the scalar potential to be $S=S\left(\vec{r}\right)$, therefore, we build a relativistic position-dependent mass system with the introduction of $S\left(\vec{r}\right)$ \cite{linear1,pdm22,pdm15,pdm16,pdm18,pdm24,vb}. In this work, we consider a scalar potential proportional to the inverse of the radial distance, then, the mass term of the Klein-Gordon equation becomes
\begin{eqnarray}
m\left(r\right)=m+\frac{\chi}{r},
\label{1.2}
\end{eqnarray}
where $\chi$ is a constant that characterizes the scalar potential $S\left(\vec{r}\right)$. It has been studied in models for confinement of quarks \cite{linear1}, in condensed matter physics \cite{cond1} and in the cosmic string spacetime \cite{eug}. In this work, we work with the Minkowski spacetime in cylindrical coordinates and the units $\hbar=c=1$:
\begin{eqnarray}
ds^{2}=-dt^{2}+dr^{2}+r^{2}\,d\varphi^{2}+dz^{2}.  
\label{1.1a}
\end{eqnarray}

Now, let us consider a background of the Lorentz symmetry violation determined by the presence of the electric field $\vec{E}=\frac{\lambda\,r}{2}\,\hat{r}$, where $\lambda$ is a constant related to a uniform volume distribution of electric charges \cite{bb20}. In this way, since the component of the Lorentz symmetry breaking tensor $\left(\kappa_{DE}\right)_{11}$ can be considered to be a constant \footnote{Note that, in this work, the component of the Lorentz symmetry breaking tensor $\left(\kappa_{DE}\right)_{11}$ is considered to be constant in the coordinate system determined by the line element (\ref{1.1a}), i.e., in cylindrical coordinates. There is nothing that forbids this assumption. However, if one changes the coordinate system, for instance, to Cartesian coordinates, then, this component of the Lorentz symmetry breaking tensor becomes a term that depends on the position. Therefore, by changing to Cartesian coordinates, it represents a soliton-like coefficient that varies with the position and breaks the momentum conservation.}, then, the Klein-Gordon equation (\ref{1.1}) becomes
\begin{eqnarray}
\left(m^{2}+\frac{2m\chi}{r}+\frac{\chi^{2}}{r^{2}}\right)\,\Phi=-\frac{\partial^{2}\Phi}{\partial t^{2}}+\frac{\partial^{2}\Phi}{\partial r^{2}}+\frac{1}{r}\frac{\partial\Phi}{\partial r}+\frac{1}{r^{2}}\frac{\partial^{2}\Phi}{\partial\varphi^{2}}+\frac{\partial^{2}\Phi}{\partial z^{2}}-\frac{g\,\left(\kappa_{DE}\right)_{11}\,\lambda^{2}\,r^{2}}{8}\,\Phi.
\label{1.4}
\end{eqnarray}

Observe that the last term of the right-hand side of the Klein-Gordon equation (\ref{1.4}) gives rise to a harmonic-type potential \cite{bb20}. Note that this system has the cylindrical symmetry, therefore, the eigenvalues of the $z$-component of the angular momentum operator and the $z$-component of the linear momentum operator are conserved quantities. This allows us to write the solution to Eq. (\ref{1.4}) in the form: $\Phi\left(t,\,r,\,\varphi,\,z\right)=e^{-i\mathcal{E}\,t}\,e^{i\,l\,\varphi}\,e^{ip_{z}\,z}\,f\left(r\right)$, where $l=0,\pm1,\pm2,\pm3,\ldots$ are the eigenvalues of the $z$-component of the angular momentum operator and $p_{z}=\mathrm{const}$ are the eigenvalues of the $z$-component of the linear momentum operator. In the following, let us define the parameters:
\begin{eqnarray}
\omega^{2}&=&\frac{g\,\left(\kappa_{DE}\right)_{11}\,\lambda^{2}}{8};\nonumber\\
\theta&=&2m\,\chi;\label{1.5}\\
\beta&=&\mathcal{E}^{2}-m^{2}-p_{z}^{2}.\nonumber
\end{eqnarray}

We proceed with a change of variables given by $\xi=\sqrt{\omega}\,r$, and thus, we rewrite Eq. (\ref{1.4}) in the form:
\begin{eqnarray}
f''+\frac{1}{\xi}\,f'-\frac{\left(l^{2}+\chi^{2}\right)}{\xi^{2}}\,f-\xi^{2}\,f-\frac{\theta}{\sqrt{\omega}\,\xi}\,f+\frac{\beta}{\omega}\,f=0.
\label{1.6}
\end{eqnarray}
By imposing that when $\xi\rightarrow\infty$ and $\xi\rightarrow0$, then $f\left(\xi\right)\rightarrow0$, therefore the function $f\left(\xi\right)$ can be written in terms of an unknown function $h\left(\xi\right)$ as follows:
\begin{eqnarray}
f\left(\xi\right)=\xi^{\sqrt{l^{2}+\chi^{2}}}\,e^{-\frac{\xi^{2}}{2}}\,h\left(\xi\right).
\label{1.7}
\end{eqnarray}

Next, by substituting Eq. (\ref{1.7}) into Eq. (\ref{1.6}), we find out that the function $h\left(\xi\right)$ is a solution to the second order differential equation:
\begin{eqnarray}
h''+\left[\frac{1+2\sqrt{l^{2}+\chi^{2}}}{\xi}-2\xi\right]h'+\left[\frac{\beta}{\omega}-2-2\sqrt{l^{2}+\chi^{2}}-\frac{\theta}{\sqrt{\omega}\,\xi}\right]=0.
\label{1.8}
\end{eqnarray}
Eq. (\ref{1.8}) is called in the literature as the biconfluent Heun equation \cite{heun,eug} and the function $h\left(\xi\right)=H_{\mathrm{B}}\left(2\sqrt{l^{2}+\chi^{2}},\,0,\,\frac{\beta}{\omega},\,\frac{2\theta}{\sqrt{\omega}};\,\xi\right)$ is the biconfluent Heun function.

Let us write $h\left(\xi\right)=\sum_{k=0}^{\infty}b_{k}\,\xi^{k}$, which means that $h\left(\xi\right)$ is written as a power series expansion around the origin \cite{arf,vbb}. Thereby, we substitute $h\left(\xi\right)=\sum_{k=0}^{\infty}b_{k}\,\xi^{k}$ into Eq. (\ref{1.8}) and obtain the relation
\begin{eqnarray}
b_{1}=\frac{\theta}{\sqrt{\omega}\left(1+2\sqrt{l^{2}+\chi^{2}}\right)}\,b_{0},
\label{1.9}
\end{eqnarray}
and the recurrence relation
\begin{eqnarray}
b_{k+2}=\frac{\theta\,b_{k+1}+\sqrt{\omega}\left(2k+2+2\sqrt{l^{2}+\chi^{2}}-\frac{\beta}{\omega}\right)\,b_{k}}{\sqrt{\omega}\left(k+2\right)\left(k+2+2\sqrt{l^{2}+\chi^{2}}\right)}.
\label{1.10}
\end{eqnarray}

We go further in search of bound state solutions, thus, we must impose that the biconfluent Heun series terminates. This occurs, from Eq. (\ref{1.10}), when 
\begin{eqnarray}
\frac{\beta}{\omega}-2-2\sqrt{l^{2}+\chi^{2}}=2n;\,\,\,\,b_{n+1}=0,
\label{1.11}
\end{eqnarray} 
where $n=1,2,3,\ldots$, which means that the biconfluent Heun series becomes a polynomial of degree $n$ when the two conditions given in Eq. (\ref{1.11}) are satisfied. From the condition $\frac{\beta}{\omega}-2-2\sqrt{l^{2}+\chi^{2}}=2n$, we obtain
\begin{eqnarray}
\mathcal{E}_{n,\,l}=\pm\sqrt{m^{2}+2\omega\left(n+\sqrt{l^{2}+\chi^{2}}+1\right)+p_{z}^{2}},
\label{1.12}
\end{eqnarray}
which corresponds to the relativistic energy levels of the system. Observe that $n$ is the quantum number related to the radial modes.

Furthermore, in order that the biconfluent Heun series terminates, we also need to analyse the condition $b_{n+1}=0$ given in Eq. (\ref{1.11}). For this purpose, let us return to the series $h\left(\xi\right)=\sum_{k=0}^{\infty}b_{k}\,\xi^{k}$ and obtain the first three terms of it. We start with $b_{0}=1$ and then, from Eqs. (\ref{1.9}) and (\ref{1.10}), we have 
\begin{eqnarray}
b_{1}&=&\frac{\theta}{\sqrt{\omega}\left(1+2\sqrt{l^{2}+\chi^{2}}\right)};\nonumber\\
[-2mm]\label{1.13}\\[-2mm]
b_{2}&=&\frac{\theta^{2}}{2\omega\left(2+2\sqrt{l^{2}+\chi^{2}}\right)\left(1+2\sqrt{l^{2}+\chi^{2}}\right)}-\frac{\vartheta}{2\left(2+2\sqrt{l^{2}+\chi^{2}}\right)},\nonumber
\end{eqnarray}
where $\vartheta=\frac{\beta}{\omega}-2-2\sqrt{l^{2}+\chi^{2}}$. Hence, by dealing with the lowest energy state of the system $\left(n=1\right)$, we have that $b_{n+1}=b_{2}=0$. From this condition, we can express the parameter $\chi$ that characterizes the Coulomb-type scalar potential in terms of the parameters of the Lorentz symmetry violation, the mass of particle and the quantum numbers $\left\{n,\,l\right\}$. This relation is obtained by taking the solutions to the following fourth degree algebraic equation for $\chi$:
\begin{eqnarray}
\chi_{1,\,l}^{4}-\left(\frac{\omega}{m^{2}}+\frac{\omega^{2}}{m^{4}}\right)\chi_{1,\,l}^{2}-\frac{\omega^{2}}{m^{4}}\left(l^{2}-\frac{1}{4}\right)=0,
\label{1.14}
\end{eqnarray}
where we have labelled $\chi_{n,\,l}$ in order to emphasize that the possible values of this parameter depend on the quantum number $\left\{n,\,l\right\}$. Thereby, the four permitted values of the parameter $\chi_{1,\,l}$, which is associated with the lowest energy state of the system, are
\begin{eqnarray}
\chi_{1,\,l}^{\left(1\right)}&=&\frac{\sqrt{2}}{2m^{2}}\,\sqrt{\omega\,m^{2}+\omega^{2}+\omega\sqrt{2\omega\,m^{2}+\omega^{2}+4m^{4}l^{2}}};\nonumber\\
\chi_{1,\,l}^{\left(2\right)}&=&-\frac{\sqrt{2}}{2m^{2}}\,\sqrt{\omega\,m^{2}+\omega^{2}+\omega\sqrt{2\omega\,m^{2}+\omega^{2}+4m^{4}l^{2}}};\nonumber\\
[-2mm]\label{1.15}\\[-2mm]
\chi_{1,\,l}^{\left(3\right)}&=&\frac{1}{2m^{2}}\,\sqrt{2\omega\,m^{2}+2\omega^{2}-2\omega\sqrt{2\omega\,m^{2}+\omega^{2}+4m^{4}l^{2}}};\nonumber\\
\chi_{1,\,l}^{\left(4\right)}&=&-\frac{1}{2m^{2}}\,\sqrt{2\omega\,m^{2}+2\omega^{2}-2\omega\sqrt{2\omega\,m^{2}+\omega^{2}+4m^{4}l^{2}}}.\nonumber
\end{eqnarray}

In this way, both conditions established in Eq. (\ref{1.11}) are satisfied, and thus we obtain a polynomial solution to the function $h\left(\xi\right)$. With this information, the expression for energy level of the lowest energy state of the system $\left(n=1\right)$ is given by
\begin{eqnarray}
\mathcal{E}_{1,\,l}=\pm\sqrt{m^{2}+\sqrt{\frac{g\,\left(\kappa_{DE}\right)_{11}\,\lambda^{2}}{2}}\left(2+\sqrt{l^{2}+\chi_{1,\,l}^{2}}\right)+p_{z}^{2}}.
\label{1.16}
\end{eqnarray}

Finally, let us use the label $\chi_{n,\,l}$ and write Eq. (\ref{1.12}) in form:
\begin{eqnarray}
\mathcal{E}_{n,\,l}=\pm\sqrt{m^{2}+\sqrt{\frac{g\,\left(\kappa_{DE}\right)_{11}\,\lambda^{2}}{2}}\left(n+\sqrt{l^{2}+\chi^{2}_{n,\,l}}+1\right)+p_{z}^{2}}.
\label{1.17}
\end{eqnarray}

From Eqs. (\ref{1.11}) to (\ref{1.17}) we can observe that the effects of the Lorentz symmetry violation modify the relativistic energy levels of the Coulomb-type scalar potential \cite{greiner,bb19}. The influence of the background of the violation of the Lorentz symmetry on the Coulomb-type interaction restricts the parameter $\chi$ to a set of permitted values in order to achieve a polynomial solution to the function $h\left(\xi\right)$. As an example, for the lowest energy state of the system, the set of the permitted values of $\chi$ has been given in Eq. (\ref{1.15}).

\section{conclusions}

We have investigated relativistic effects on a scalar particle under the influence of a scalar potential proportional to the inverse of the radial distance and the violation of the Lorentz symmetry. We have considered a background of the violation of the Lorentz symmetry that yields a harmonic-type potential on the Klein-Gordon equation. Then, in search of polynomial solutions to the function $h\left(\xi\right)$ given in Eq. (\ref{1.7}), we have seen that the effects of the Lorentz symmetry violation modify the relativistic energy levels of the Coulomb-type scalar potential. The influence of the Lorentz symmetry violation background on the Coulomb-type interaction restricts the parameter $\chi$ to a set of permitted values that yield a polynomial solution to $h\left(\xi\right)$. As an example, we have obtained the permitted values of $\chi$ for the lowest energy state of the system.

\acknowledgments{The authors would like to thank the Brazilian agencies CNPq and CAPES for financial support.}

\end{document}